\documentstyle[12pt]{article}
\begin{document}
\begin{flushright}
NSF-ITP-92-115\\
TPI-MINN-92/52-T\\
September 1992\\
\end{flushright}
\begin{center}
{\bf The Rule of Discarding $1/N_c$ in Inclusive Weak Decays. II}
\end{center}
\vskip0.5in
\begin{center}
{\bf B. Blok} \\
Institute for Theoretical Physics\\
University of California, Santa Barbara, CA 93106\\
and\\
{\bf M. Shifman} \\
Theoretical Physics Institute\\
University of Minnesota,
 Minneapolis, MN 55455\\
\end{center}

\begin{center}
\vspace{0.2cm}
{\bf\large Abstract}
\end{center}
{We calculate the leading preasymptotic correction to the inclusive
width $b\rightarrow \bar c c s$ (two massive
quarks in the final state) due to the operator $\vec \sigma \vec H$.
It is found that this correction tends to cancel the $1/N_c$
part of the inclusive width calculated using naive factorization.
The absolute value of the effect is of order 0.25.
We also discuss how  quark-hadron duality is realized in the inclusive
decays.
}
\vspace{0.2cm}
{
}
\vfill
\section{Introduction}
\par  It was discovered recently \cite{BUV,BS1}
 that the dominant preasymptotic
correction in the inclusive weak nonleptonic decays of heavy flavors is
due to the operator of the type $\vec\sigma\vec H$ \cite{BS2}. It is of
order $m_Q^{-2}$ relative to the asymptotic  (parton) expression
($m_Q$ is the mass of the decaying heavy quark) and occurs
in the interference part of the amplitude. The sign of this correction
is such that generically it tends to suppress the parton result
for the interference part.
\par In this note we further investigate the effects of the soft-gluon
emission
 reducible to the operator $\vec\sigma\vec H$. Two issues are considered:
the transitions of the type $b\rightarrow c\bar c s$, with two
massive quarks in the final state, and the distribution in the invariant mass
of the final quark-antiquark pair. The analysis of the latter distribution
allows us to trace, at least at the qualitative level,
the origin of the leading preasymptotic correction in the interference
part in terms of the exclusive channels \cite{BS1,BS2}, and relate it
to the rule of
 discarding $1/N_c$ in the weak exclusive decays \cite{BSW,BGR}.
\par All notations and conventions are the same as in the first part of this
work \cite{BS1}, hereafter referred to as I. The reader can also find in I
an introduction to the preasymptotic corrections  in weak
 inclusive decays and a representative list
of references.

\section{The leading preasymptotic correction in $b\rightarrow c\bar cs$}
As it was explained in I, the effective weak Lagrangian we work with has the
generic form
\begin{equation}
L_{eff}={G_F\over \sqrt{2}} V_1V_2\{ c_1(\bar B\Gamma_\mu A)
 (\bar \beta\Gamma^\mu\alpha )+c_2(\bar B_i\Gamma_\mu A^j)
 (\bar \beta_j\Gamma^\mu\alpha^i )\}.
\label{eq:1}
\end{equation}
The constant $G_F$
in eq. (\ref{eq:1}) is the Fermi coupling constant, $V_{1,2}$ are the relevant
CKM matrix elements, $c_{1,2}$ will be treated as numerical coefficients,
$\Gamma_{\mu}=\gamma_{\mu}(1+\gamma_5)$. In the transition at hand we
identify
\begin{equation}
A=b,\,\,B=c,\,\,\beta=s,\,\,\alpha =c.
\label{eq:2}
\end{equation}
It is convenient to write the operators $O_1$ and $O_2$ in the
following form:
\begin{equation}
O_2 = (\bar \beta\Gamma_\mu A)
 (\bar B\Gamma_\mu\alpha )\equiv (\bar s\Gamma_\mu b)(\bar c\Gamma_\mu c),
\label{eq:3a}
\end{equation}
\begin{equation}
 O_1={1\over N_c}O_2+\tilde O_1,
\label{eq:3b}
\end{equation}
\begin{equation}
\tilde O_1=2
(\bar \beta\Gamma_\mu t^a A)
 (\bar B\Gamma_\mu t^a\alpha )\equiv 2 (\bar s\Gamma_\mu t^ab)(\bar c\Gamma_\mu
 t^ac),\,\, ({\rm Tr}t^at^b=(1/2)\delta^{ab}).
\label{eq:3c}
\end{equation}
In the parton model the interference part of $\Gamma( b\rightarrow c\bar c s)$
is well known in the literature (see e.g. ref.
\cite{XX}).
\begin{equation}
\Gamma^{(0)}_{\rm int}={1\over m_{H_A\bar\gamma}}
G^2_F\mid V_1V_2\mid^2c_1c_2{\rm Im}
<H_{A\bar\gamma}\mid \int d^4xiT\{O_2(x)
 O_1^+(0)\}\mid H_{A\bar\gamma}>
\label{eq:4}
\end{equation}
$$={G^2_F\over 96\pi^3}\mid V_1 V_2\mid^2 c_1c_2 M^5F_3(m^2/M^2)$$
where m is the mass of the c quarks while M is that of b. The subscript
"int" marks  the interference part of the amplitude.
The function $F_3$ reflects the effect of the nonvanishing c-quark mass --
it would be 1 if m=0.
Actually,
\begin{equation}
F_3(x)=v_0(1-14x-2x^2-12x^3)+24x^2(1-x^2){\rm ln}((1+v_0)/ (1-v_0)),
\label{eq:5}
\end{equation}
$$
v_0=\sqrt{1-4x},\quad x=m^2/M^2.
$$
Observe that, in accordance with the general rule, $\Gamma_{\rm int}\sim N_c^0$
while the noninterference part is proportional to $N_c^1$.
\par The asymptotic (parton) expression for $\Gamma_{\rm int}$ given in
eq. (\ref{eq:4})
 obviously corresponds to the correlation function of
$O_2$ with the trivial component of $O_1^+$, namely that where the
$\bar c c$ pair is in the color singlet state (see the
decomposition in eq. (\ref{eq:3b}); in the parton model only the term $O_2/N_c$
is relevant). The leading preasymptotic correction is  due to the second
component of $O_1^+$, namely $\tilde O_1$, with the $\bar c c$
pair in the color-octet
 state. This implies a soft gluon emission giving rise to the operator
$\bar b \Gamma^\mu G^a_{\alpha\mu}t^ab$, see Fig.1.
 Calculation of the coefficient in front of the operator
$\bar b\Gamma^\mu G_{\alpha\mu}^at^ab$
 is rather straightforward, we comment here only on a few
basic points. Some details relevant to the discussion of the invariant mass
distribution are given in  Appendix.
\par Technically it is very convenient to carry out all
 computations using the background
field method ( a pedagogical review is presented in ref. \cite{N}).
We expand the quark Green functions in the background field keeping
only the term $O(G^0)$ and $O(G^1)$ where G is a generic notation for the
gluon field strength tensor.
 Terms of higher order in G or terms with derivatives are irrelevant
to the operator at hand.
The  quark propagator in the coordinate space has the form
\begin{equation}
S(x,0)={1\over 2\pi^2}{\hat x\over x^4}\{-{1\over 2}
m^2x^2K_2(m\sqrt{-x^2})\}
-{1\over 8\pi^2}{x^\alpha\tilde G_{\alpha\varphi}\gamma^\varphi\gamma^5
\over x^2}
\label{eq:6}
\end{equation}
$$
\{{-x^2mK_1(m\sqrt{-x^2})\over \sqrt{-x^2}}\}+({\rm
terms}\,\, {\rm with}\,\, {\rm the}
\,\, {\rm even} \,\, {\rm number} \,\, {\rm of}\,\, \gamma\,\,{\rm matrices}
).
$$
We use  the Schwinger gauge condition on the background field:
\begin{equation}
x_\mu A^\mu (x)=0,\quad A_\mu\equiv g A^a_\mu t^a.
\label{eq:61}
\end{equation}
 The expressions in braces account for the quark mass; they reduce
to 1 in the limit $m\rightarrow 0$; $K_{1,2}$ stand for the
 corresponding McDonald functions.
 The terms in $S(x,0)$ containing even number of $\gamma$ matrices
drop out because of the chiral structure of the operators $O_{1,2}$
( $\Gamma'$s in the vertices ensure that the only relevant
part of $S(x,0)$ is the one with the odd number of $\gamma$ matrices).
\par Now, using the general structure of the Green function
it is very easy to show that the gluon emission occurs
only from the antiquark line (Fig.1). The emission from the c and s quark
lines vanishes. This property could have been foreseen on
general grounds. Indeed, the operators $O_{1,2}$ written in the form
(\ref{eq:3b}), force the gluon emission
 inside the $c\bar c$ loop. On the other hand,
one could Fierz rearrange these operators prior to explicit computation.
Then one will
 conclude that the gluon emission must occur inside the $\bar c s$ loop.
\par The actual calculation is carried out in two stages. The correlation
function
\begin{equation}
\Gamma^{(1)}_{\rm int}={1\over m_{H_A\bar\gamma}}
G^2_F\mid V_1V_2\mid^2c_1c_2 {\rm Im}<H_{A\bar\gamma}\mid \int d^4xiT\{O_2(x)
\tilde O_1^+(0)\}\mid H_{A\bar\gamma}>
\label{eq:7}
\end{equation}
contains as a natural block the loop graph of Fig. 2
where the
 thick line denotes the c-quark Green function in the background gluon field,
one of the vertices is $\Gamma_\mu t^a$ while the other one is $\Gamma_\nu$.
 Since we are interested only in  terms linear in
$G_{\mu\nu}^a$ the position of the color matrix $t^a$ is unimportant;
one can move it freely along the thick line.
Let us denote by $ q$ is the total momentum of the $c\bar c$ pair.
 Once the imaginary part of the loop in Fig. 2
is found as a function of $q^2$
it is quite trivial to determine the complete
imaginary part in eq. (\ref{eq:7}) simply by integrating over the
phase space of the s quark.
\par After these explanatory remarks let proceed directly to the results. The
 imaginary part of the Feynman graph on Fig. 2 is
\begin{equation}
{\rm Im} D^a_{\mu\nu}={1\over 8\pi}{1-v^2\over sv}\{\tilde G^a_{\mu\alpha}
q^\alpha
 q_\nu
+(\mu\leftrightarrow \nu)\}\label{eq:8}
\end{equation}
where\begin{equation}
s=q^2,\quad v\equiv (1-4m^2/ s)^{1/2},\label{eq:9}
\end{equation}
and $D_{\mu\nu}^a$ is defined on Fig. 2.
Notice that in the limit $m\rightarrow 0$ the coefficient function in front of
the braces in eq. (\ref{eq:8}) is actually $\delta (s)$; we will return
to discussion of this point later.
\par Furthermore, the imaginary part in eq. (\ref{eq:7}) reduces to that
of $D^a_{\mu\nu}$,
\begin{equation}
{\rm Im}\,  iT\{\int d^4x O_2(x)\tilde O_1^+(0)\}
=2\int ({\rm Im} D_{\mu\nu}^a)
{d^3\vec p\over 2p^0(2\pi)^3}(\bar b t^a\Gamma^\mu\hat p\Gamma^\nu b)
\label{eq:10}
\end{equation}
where $p=(p^0,\vec p)$ is the on-mass-shell momentum of the
s quark and the factor of 2 is the same factor that appears in the
definition of the operator $\tilde O_1$
(see eq. (\ref{eq:3c})).
\par Integration over the phase space is readily performed,
and we arrive at the following expression for
$\Gamma_{\rm int}^{(1)}$:
\begin{equation}
\Gamma^{(1)}_{\rm int}=-{G^2_F\mid V_1V_2\mid^2c_1c_2\over m_{H_{A\bar\gamma}}}
{M^3\over 24\pi^3}
<H_{A\bar\gamma}\mid \bar A \tilde G_{0i}
\gamma^i\gamma^5 A\mid H_{A\bar\gamma}>F_4(m^2/M^2)
\label{eq:11}
\end{equation}
where
\begin{equation}
F_4(x)=v_0(1+(1/2) x +3x^2)-3x(1-2x^2)\ln{{(1+v_0)^2\over 4x}}
\label{eq:12}
\end{equation}
and the notations are the same as in eq. (\ref{eq:5}).
In passing to eq. (\ref{eq:11}) we have restored the generic notations for the
quarks and mesons, as in I. For instance
$H_{A\bar\gamma}$ denotes the meson with the quark content
$A\bar \gamma$, etc. It is worth reminding that A is the same as the b quark
in the problem at hand.
\par The function $F_4$ reflects
 the masses of c and $\bar c$ in the final state. If m=0 $F_4(x=0)=1$
and we reproduce the expression for $\Delta \Gamma_{\rm hadr}$
from refs. \cite{BUV,BS1}. As explained in ref. \cite{BS2}
the hadronic matrix element in eq. (\ref{eq:11})
(it is written here in the A rest frame, assuming M$\rightarrow \infty$) is
expressible in terms of the observable quantities
\begin{equation}
<H_{A\bar\gamma}\mid \bar A\tilde G_{0i}\gamma^i\gamma^5A\mid H_{A\bar\gamma}>
=2m^2_{\sigma H}M.
              \label{eq:13}
\end{equation}
Here $m_{\sigma H}^2=(3/4)(m_{B^*}^2-m_B^2)$.
Combining eqs. (\ref{eq:4}) and (\ref{eq:13})
we obtain
\begin{equation}
{\Gamma^{(1)}_{\rm int}\over \Gamma ^{(0)}_{\rm int}}
= -8\frac{m^2_{\sigma H}}{M^2}
{F_4(x)\over F_3(x)}.\label{eq:14}
\end{equation}
At this point we pause to mention a subtlety associated with the parameter M
$-$ the quark mass. Unlike $m^2_{\sigma H}$ this parameter is not
directly observable and, hence, requires comments. A constant shift
in $M$, i.e. independent of the heavy quark
mass, gives rise to a power correction analogous to
the one in eq. (\ref{eq:14}),
 generally speaking. In other words one can say that redefining $M^5$ in eq.
(\ref{eq:4}) one generates a universal power correction in the non-leptonic
 widths of all $A$-flavored hadrons.
\par Since we are interested only in the leading power correction,
 the definition of the quark mass
will not change $\Gamma_{\rm int}^{(1)}$.
This is not the case in $\Gamma^{(0)}_{\rm int}$, however.
 As a matter of fact, it is not difficult to see that the
bare graph corresponding to $\Gamma^{(0)}_{\rm int}$ (analog of Fig. 1,
but  with
no soft gluon emission from the $c\bar c$ loop)
is proportional to $P^4\hat{P}$ where $P$
 is the momentum of the
initial quark $A$. In terms of the covariant quantities one can identically
rewrite \cite{BUV} (see also Appendix)
\begin{equation}
\bar A P^4\hat PA=\bar A\{\hat{\cal P}^5-{i\over 2}((G
_{\alpha\beta}\sigma^{\alpha\beta})\hat{\cal P}^3+\hat{\cal P}^3
(G
_{\alpha\beta}\sigma^{\alpha\beta}))\}A
\label{eq:15}
\end{equation}
where ${\cal P}$ is the covariant momentum operator, ${\cal P}_\mu=iD_\mu$.
It is instructive to notice that in the limit $M\rightarrow \infty$
the operator $P^\alpha\bar A\tilde G_{\alpha\beta}\Gamma^\beta A$ appearing in
eq. (\ref{eq:11}) and the operator $\bar AG_{\alpha\beta}\sigma^{\alpha\beta}
A$ in eq. (\ref{eq:15})
reduce to each other. Namely, in the nonrelativistic limit
\begin{equation}
\bar A\tilde G_{0i}\gamma^i\gamma^5A\rightarrow -\bar h_A\vec \sigma
 \vec H h_A,
\,\,
\bar A G_{\alpha\beta}\sigma^{\alpha\beta}A
\rightarrow 2i\bar h_A\vec \sigma \vec H h_A.
\label{eq:15789}
\end{equation}
\par Eq. (\ref{eq:15}) implies (through
equations of motion) that $M^5$ in eq. (\ref{eq:4}) is actually to be
understood
as $M_0^5-2M_0^3m^2_{\sigma H}$ where $M_0$ is the "bare" quark mass
entering equations of motion.
\par In the above arguments the Schwinger gauge condition on the background
gluon field is always implied. Had
we chosen another gauge for the background field additional graphs
with the soft gluon line attached would appear. The final assertions are,
of course, gauge independent.
\par Concluding the section let us present a few numerical estimates.
We assume the $b$ quark mass $M$ to be  equal to 4.5 GeV and the charmed
quark mass $m$ to 1.35 GeV.
Then $x\sim 0.1$, $F_3=0.15$, $F_4=0.23$.
For the ratio of the factorizable and nonfactorizable parts
of the $1/N_c$ piece of the amplitude we obtain
\begin{equation}
r={\Gamma_{\rm int}^{(1)}\over \Gamma_{\rm int}^{(0)}}
= -8{m^2_{\sigma H}\over m_B^2}
{F_4(0.1)\over F_3(0.1)}\sim -0.25.
\label{eq:1617}
\end{equation}
It is  seen that this  ratio is numerically the same as
the corresponding ratio in
the $b\rightarrow c\bar u d $ channel, as it  was calculated in refs.
\cite{BUV,BS1}.

\section{Duality. The invariant mass distributions}
\par This section is devoted to related questions: (i) how the interference
part
of the inclusive width might be saturated by specific hadronic states;
(ii) duality in the asymptotic regime and preasymptotic corrections.
\par We will check the self-consistency of our approach. First,
we will see that quark-hadron duality is valid at each step of our
calculations. Second, we will demonstrate
that the sum of exclusive channels nicely matches the
inclusive width.
\par The discussion of  how duality might be  implemented
in $\Gamma_{\rm int}$ was started in \cite{S} (see also I). Here
we present additional
 arguments which seemingly solve the problem at the qualitative level.
The invariant mass distribution, to be considered below, has a remarkable
 structure,
with far-going consequences.
\par Let us recall that in our calculation of the nonfactorizable
contributions to
the exclusive widths we used the expansion in the parameters (see
\cite{BS2})
\begin{equation}
\lambda\sim {\Delta\mu\over Q^2},\quad\lambda'={\mu^2\over Q^2}
\label{eq:1511}
\end{equation}
where $\Delta$ is the energy release,
$\mu$ is a typical off-shellness
of quarks (gluons) inside hadrons and
$Q^2\sim 1$ GeV$^2$ can be considered roughly speaking
as the invariant mass of the light quark pair forming the
hadronic state at hand.
If $\lambda'\ll 1$ and the operator product expansion is applicable we have
two possible kinematical regimes for $\lambda$:
$$
1)\quad\lambda\le 1,
$$

$$2)\quad\lambda\ge 1.$$
\par The first regime corresponds to exclusive decays
analysed in \cite{BS2} while the second one corresponds to the
inclusive description. Generally speaking,  situation
with the power corrections in these two domains is different,
and one should always keep this difference in mind.
Still, one might hope to get a sort of matching, and we analyse
this possibility. In other words, it will be explicitly
 checked how our predictions for exclusive and inclusive widths match  each
other.
\par Since the issue is quite general, to begin with it is convenient to
 get rid
of inessential details and choose kinematics in such a way as to make all
assertions most transparent.
\par Thus, we consider a generic process
$A\rightarrow B\beta\bar\alpha$.
We consider the kinematical regime $\lambda\geq 1$, relevant to inclusive
processes and try to understand how the inclusive rate
is saturated by exclusive channels.
For simplicity we impose an  additional constraint
(which helps technically but is otherwise unimportant),
\begin{equation}
M_A,M_B\rightarrow\infty , M_A-M_B=\Delta\,\, {\rm fixed},\,\,  m_\alpha =
m_\beta =0.
\label{eq:1514}
\end{equation}
This means that we to work in the small velocity (SV) limit \cite{VS}, avoiding
complications due to recoil dependence of various formfactors.
\par The central idea is as follows. In the asymptotic limit
where the preasymptotic corrections are negligible there exist a natural
line of reasoning  showing that neither the naive factorization
nor the rule of discarding $1/N_c$ can
take place universally (see I). The actual situation is more complicated.
Presumably, for the
 operator
$O_1=(\bar B\Gamma_\mu A)(
\bar\beta\Gamma_\mu\alpha)$
the rule of discarding $1/N_c$ holds maximally while for the operator
$O_2^+=(\bar A\Gamma_\mu\beta)(\bar\alpha\Gamma^\mu B)$
it is maximally violated, and the naive factorization takes place.
\footnote{We hasten to add that this asymptotic "inclusive" picture is
quite opposite to what happens in the exclusive decay
$B^0\rightarrow D^+\pi^-$. It has been shown \cite{BS2}
that the rule of discarding $1/N_c$ is applicable to $O_2$
in this decay. The breakdown of the expected asymptotic picture
in the exclusive channels is correlated with the preasymptotic corrections
to the inclusive rates, as will be discussed below.}
The arguments substantiating the point are
admittedly heuristic and are based on the notion of color transparency
(see e.g. \cite{B}).
\par Let us elucidate the above assertion in more detail. We consider
now the limit of infinitely large
$M_{A,B}$ with the preasymptotic corrections switched off.
(The parton-model limit). As explained in I the absorptive part of
the interference amplitude is contributed, in the
general case, by the intermediate
hadronic states of 3 distinct types.
\par (i) The operator $O_1=(\bar B\Gamma_\mu A)(
\bar\beta\Gamma^\mu\alpha)$ produces the state $\{B\bar\gamma\}+\{\beta
\bar\alpha\}$ in the leading order in $1/N_c$; this state is absorbed by the
 operator $O_2^+=(\bar A\Gamma_\mu\beta)(\bar\alpha\Gamma^\mu B)$
at the level $1/N_c$ (the figure bracket denotes a color-singlet meson state,
possibly excited, with the given quark content);
\par (ii) the operator $O_1$ produces the state
 $
\{B\bar\alpha\}+
\{\beta\bar\gamma\}$
at the level $1/N_c$ which is then swallowed by $O_2^+$ in the leading order.
\par (iii) the operator $O_1$ produces the state $\{\beta\bar\alpha\}+
\{B\bar\beta\}+
\{\beta\bar\gamma\}$ at the level $1/\sqrt{N_c}$;
 this
 state is then annihilated by $O_2^+$ at
the level
$1/\sqrt{N_c}$.
\par Let us argue that only the first option is realized in the kinematic
 domain
considered. The state
$
\{B\bar\alpha\}+
\{\beta\bar\gamma\}$
that formally has a projection on $O_1$ of order $1/N_c$ actually has an
 intrinsic suppression
by powers of $\Lambda_{QCD}/\Delta$, where $\Delta$ is the energy release.
($\Delta=M_A-M_B$). Indeed, in the SV point ($\vec v_B\rightarrow 0
$) the quark B "gently" substitutes A.  The $\{\bar\gamma B\}$ system
remains undistorted. The $\bar\alpha\beta$ pair produced at the time t=0 does
not interact with the surrounding medium till $t\sim \Lambda_{QCD}^{-1}$
in its rest frame, i.e. till $t\sim \Delta/\Lambda_{QCD}^2$
 in the rest frame of A.
By that time the $\bar \alpha\beta $  system is separated from $B\bar \gamma$
by distance of order $\Delta/\Lambda_{QCD}^2\gg\Lambda_{QCD}^{-1}.$
{}From the point of view of the $\{\beta\bar\alpha\}$
system at this moment of time the state $\{B\bar
\gamma\}$ presents a tiny white object far beyond the correlation length.
Therefore, it disappears "below" the horizon and decouples
(Fig. 3). The color
twisted state $\{B\bar\alpha\}+\{\beta\bar\gamma\}$ does not evolve from
$O_1$. The corresponding suppression is governed not only by $1/N_c$;
there is another parameter,
 $\Lambda_{QCD}/\Delta$.
 \par One can identically rewrite $O_1$ using the Fierz transformation,
\begin{equation}
O_1={1\over N_c}(\bar\beta\Gamma_\mu A)(\bar B\Gamma^\mu\alpha)
+2(\bar\beta\Gamma_\mu t^aA)(\bar B\Gamma^\mu t^a\alpha).
\end{equation}
This means that at short times the operator may $O_1$ form a superposition
of two states of the type
 $"\beta\bar\gamma"_s+"B\bar\alpha"_s$
plus
 $"\beta\bar\gamma"_o+"B\bar\alpha"_o$
(where the subscripts $s$ and $o$
mark color singlet and octet), but at large times the  octet $\times$
octet part cancels singlet $\times$ singlet , and the rule of discarding
 $1/N_c$
holds for the production of color-twisted states by $O_1$ (Fig. 4).
In other words, the states of the type (ii) do not play a role in the
saturation of $\Gamma^{(0)}_{\rm int}$.
\par Proceed now to the states of the type (iii). Here the situation
 seems to be the simplest.
The production of the extra $\beta\bar\beta$ pair from $\{B\bar\gamma\}
$ is suppressed by  $1/\sqrt{N_c}$ and $\mid\vec v_B\mid$, where
$\mid\vec v_B\mid$
is the B-quark velocity in the A rest frame. Hence in the SV limit
 this contribution
is absent.
\par The last point to be discussed is the states of the type (i) and
their projection
on $O_2^+$.  The operator $O_2$
after the Fierz rearrangement takes the form
\begin{equation}
O_2={1\over N_c}(\bar B\Gamma_\mu A)(\bar \beta\Gamma^\mu\alpha)+2(\bar B
\Gamma_\mu t^a A)(\bar \beta \Gamma^\mu t^a\alpha )
\label{eq:Fierz}
\end{equation}
It produces the desired state at the level $1/N_c$ plus the octet
$\times$ octet component (Fig. 5).
The latter includes
 a pair of fast light quarks. This latter component asymptotically
vanishes, presumably due to the strong radiation of gluons (the Sudakov
effect).
Thus, for the color twisted states produced by $O_2$ we have exact
naive factorization at the level of $1/N_c$.
(Warning: we speak here of highly excited states
typical of the inclusive production, see the footnote on page 8).
As a result of this asymmetry
in the hadronic saturation (the rule of discarding $1/N_c$ for $O_1$
and the naive factorization for $O_2$) the duality in the interference part
of the amplitude  is now perfectly satisfied. The origin
of the asymmetry between $O_1$ and $O_2$ is specific kinematics, of course:
the SV limit for B and large energy release --
fast $\beta\bar\alpha$. Under different kinematic conditions
the pattern of saturation can be different \cite{S}.
\par The naive factorization for the operator $O_2$ that we advocate
 here is of course not in contradiction
with the results of ref. \cite{BS2}, where the rule of discarding
$1/N_c$ has been shown to emerge for the same operator, but
in the different kinematic regime.
In ref. \cite{BS2} the light quark pair we dealt with had a limited invariant
 mass, a few units times $\Lambda_{QCD}$.
In the inclusive decays in the parton-model regime discussed above
the distribution over the $\beta\bar\alpha$ invariant mass is of order
$\Delta$, and this parameter is assumed to be parametrically much larger than
$\Lambda_{QCD}$. Therefore,  the rule of discarding $1/N_c$
for small invariant masses and the naive factorization for large
 invariant masses
for one and the same operator may well coexist. Shortly, when we
proceed to the discussion
 of the
preasymptotic corrections, we will find additional arguments supporting this
assertion.
\par We saw above how duality is realized in $\Gamma^{(0)}_{\rm int}$.
Let us now consider the preasymptotic corrections.
\par Now we come to the most essential question$-$the
 invariant mass distribution
for the light quark pair ($\beta\bar\alpha$), produced by the operator
$O_1$ and absorbed by $O_2$. The point is that in the case of the soft gluon
emission  (operator $\vec\sigma\vec H$) this distribution is drastically
different from the one that  occurs
 in the parton part of the amplitude. Indeed, let us
analyse the expression for ${\rm Im}
 D^a_{\mu\nu}$, given by eq. (\ref{eq:8}) in the
limit $m\rightarrow 0$. Recall that s is the invariant mass squared
of the $\beta\bar\alpha$ pair. For fixed s and $m\rightarrow 0$ the parameter v
defined
 in eq. (\ref{eq:9}) tends to 1 and formally ${\rm Im} D^a_{\mu\nu}$ vanishes
(see eq. (\ref{eq:8})).
It is not difficult to understand, however, that the above assertion
 of vanishing takes
place only at $s\ne 0$, and as a matter of fact at $m\rightarrow 0$
${\rm Im} D_{\mu\nu}^a$ is proportional to $\delta (s)$. Specifically,
\begin{equation}
{\rm Im} D^a_{\mu\nu}\mid _{m\rightarrow 0}={1\over 4\pi}\delta (s)
(\tilde G_{\mu\alpha}q_\alpha q_\mu +(\mu \leftrightarrow \nu )).
\label{eq:16}
\end{equation}
(Note the similarity of this expression to the Dolgov-Zakharov-'t Hooft
approach
to the triangle anomaly \cite{DZ,T}).
Eq. (\ref{eq:16}) has a natural interpretation in terms of hadronic states.
It implies that the hadronic
 states responsible for the leading preasymptotic correction
 are characterized by a small invariant mass
of the $\beta\bar\alpha$ pair. Then, these quarks have to materialize
in the form of {\em light} mesons, like $\pi$
 and $\rho$.
In other words, the Dalitz plot, which is relatively homogeneously populated
in the parton model is distorted in the domain corresponding to small
 invariant masses
of $\beta\bar\alpha$ pair. The distortion corresponds to an extra negative
contribution relative to that emerging in the factorized approximation.
The operator $O_2$ which obeys the naive factorization for highly excited
states (populating the interior of the Dalitz plot) switches
to the rule of discarding $1/N_c$ for low invariant masses of
$\beta\bar\alpha$ (i.e. at the boundary of the Dalitz plot).
\par Indeed, if we consider the interference part of the amplitude,
in the factorized approximation in the low $s$ domain, the contribution
of $D\pi, \,\, D\rho$, etc. is
 obviously proportional to ${f^2_\pi\over N_c} \Delta^3$. At the same time
the preasymptotic correction due to $\vec \sigma\vec H$ is
 $\sim m^2_{\sigma H} \Delta^3$. The ratio of these two effects is
 $N_cm^2_{\sigma H}/f^2_\pi $, i.e. exactly the same as has been found in
 ref. \cite{BS2}
for exclusive decays of the type $B\rightarrow D\pi$,
$B\rightarrow D\rho$.
Thus, the correction discussed in I and in the present paper can be most
 probably
associated with the distortion of the decay modes where $\beta\bar\alpha$
pair forms generic $\pi'$s and $\rho'$s -- dynamical emergence of the rule of
discarding $1/N_c$ in the exclusive decays of this kind.
\par It is instructive to trace not only the parametric dependence like
$N_cm^2_{\sigma H}/f^2_\pi$ mentioned above -- which comes out correct
-- but to work out numbers as well. The contribution of $ D\pi$
 and $ D^*\pi$ in
the factorization approximation is (neglecting the pion mass):
\begin{equation}
\Delta\Gamma_{\rm fact}=m^3_B(1-{m^2_D\over m^2_B})^3
{1\over 16\pi}\frac{f^2_\pi}{N_c}\times 2.
\label{eq:18}
\end{equation}
Here and below we measure $\Delta\Gamma$ in the units of
$G_F^2|V_1V_2|^2c_1c_2$.
The Isgur-Wise function $\xi =1$
in the limit of eq. (\ref{eq:1514}).
Moreover, if we neglect $m_\rho$ compared to $\Delta$
the contribution  of $ D\rho$ and $ D^*\rho$ is the same as above, with an
obvious substitution $f_\pi^2\rightarrow f_\rho^2$.
According to ref. \cite{BS2} the nonfactorizable
part of the same contribution for  $D\pi$ and $D^*\pi$  is
\begin{equation}
\Delta\Gamma_{\rm n.f.}=-m^3_B(1-{m^2_D\over m^2_B})^3
{1\over 64\pi^3}m^2_{\sigma H}\times (1+\frac{1}{3}).
\label{eq:18a}
\end{equation}
Here the factor 4/3 takes into account  different degrees of cancellation
of the $1/N_c$ parts in $D\pi$ and $D^*\pi$; we also omit all factors like
$x$ or $x'$ which are of order one but  are not exactly fixed by
our calculation
(see ref. \cite{BS2} ).
For the  case of the decay with the production  of
$\rho$ the answer is essentially the same, up to corrections
due to the nonvanishing  mass of the $\rho$ meson.
Neglecting this
mass  we just double the result. Thus, we get, summing
the contributions of these
four channels in the nonfactorizable part of $\Gamma_{\rm int}$:
\begin{equation}
\Delta\Gamma_{\rm n.f.}=-m^3_B(1-{m^2_D\over m^2_B})^3
{1\over 24\pi^3}m^2_{\sigma H}.
\label{eq:19}
\end{equation}
This result is remarkably close to
 the nonfactorizable contribution in the inclusive
width $\Gamma_{\rm int}^{(1)} $ calculated in \cite{BUV,BS1}. As a matter
of fact, eq. (26) in I gives for $\Gamma_{\rm int}^{(1)} $
the same expression with the substitution $(1/24\pi^3)
\rightarrow (1/12\pi^3)$. Thus, if we stick literally to our formulae in the
exclusive case we find that the nonfactorizable correction in the
four exclusive channels is responsible for $\sim 0.5$ of the
inclusive nonfactorizable correction. The remaining 1/2 may be due to, say,
$a_1$ contribution. It may well be, however, that various uncertainties of
order one in our exclusive estimates actually add up to increase
 the nonfactorizable parts of some of these four exclusive amplitudes,
so that they  build up the  inclusive correction (almost) entirely.
\par Summarizing, one can say
 that the nonfactorizable correction due to the operator
$\bar b\tilde G^a_{0i}t^a\gamma^i\gamma^5 b$
 in the inclusive width nicely matches the corresponding
sum in the exclusive channels.
Quark-hadron duality is valid at each
step of our calculation.
\section{Conclusions}
\par In this paper we found that the leading preasymptotic correction
to the inclusive width in the channel $b\rightarrow \bar c c s$
cancels about 25 $\%$ of the $1/N_c$ suppressed part of the inclusive width
calculated in the naive parton model. As in I, this effect leads
 to an additional increase of the inclusive
hadronic width in this channel by $\sim 5\%$.
It would be interesting to have a direct experimental measurement of this
width.
\par We have also shown that the analysis of the invariant mass
distribution allows one, at least at the qualitative level,
to trace the way quark-hadron duality is realised in the weak decays
 of heavy quarks, and to understand the relation between
inclusive and exclusive widths. In particular, we have demonstrated,
as it could be expected already from the results of I
that the leading preasymptotic correction to the inclusive width of B
meson is saturated by the exclusive widths of the type
$B\rightarrow D$ + light meson, $B\rightarrow D^*$ + light meson,
where the set of light mesons seems to be exhausted by
$\pi$, $\rho$ and may be $A_1$ and $\rho'$.
\par It would be of interest to calculate next-to-leading
preasymptotic corrections both for  $b\rightarrow c\bar ud$ and
$b\rightarrow c \bar cs$ channels and to study
further the pertubative QCD corrections \cite{L}.
\par The authors are grateful to A. Dobrovolskaya, A. Vainshtein
and M. Whitterell for
useful discussions.

\section{Appendix. Calculations in External Fields in the Interference
Part of the Amplitude}
\par In this appendix we provide a few useful computational
details relevant to the calculation of
${\rm Im} D_{\mu\nu}^a$, $\Gamma^{(1)}_{\rm int}$
, and the covariantization of the parton model calculation
of $\Gamma^{(0)}_{\rm int}$.
\par We consider first the c-quark loop in the external field.
This loop can be formally written
as
\begin{equation}
D^a_{\mu\nu}=i{\rm Tr}\,\{\Gamma_\mu t^a{
1\over \hat{\cal P}-\hat q/2 -m}\Gamma_\nu{1\over \hat{\cal P}+\hat q/2 -m}\}
\label{eq:A.1}
\end{equation}
Here $\hat{\cal P}$ is the formal covariant momentum operator
(see ref. \cite{N}). Since we are interested only in terms linear in
$G^a_{\mu\nu}$, the gluon field strength tensor, we can assume that
$G^a_{\mu\nu}$ is constant ($x$-independent) c-number field. Using the
general properties of the trace operation and the fact that in our
approximation $t^a$ can be moved to any position freely, it is trivial to
show that $D^a_{\mu\nu}=D^a_{\nu\mu}$.
\par Now, we can use the fact that
\begin {equation}
(\hat{\cal P} -m)^{-1}
=(\hat{\cal P} +m)({\cal P}^2-m^2+{i\over 2}\sigma_{\alpha\beta}
G_{\alpha\beta})^{-1}.
\label{eq:A.2}
\end{equation}
The term with $\sigma_{\alpha\beta}G_{\alpha\beta}$ is the magnetic
emission, and this source of $G^a_{\mu\nu}$ is quite obvious. One should
notice, though, that $G^a_{\mu\nu}$ is also hidden in ${\cal P}_\alpha$; this
is
the electric emission.
Using eq. (\ref{eq:A.2}) we rewrite eq. (\ref{eq:A.1}) in the following form
\begin{equation}
D^a_{\mu\nu}=i{\rm Tr}\,\{\Gamma_\mu t^a(\hat{\cal P}-\hat q/2){1\over
({\cal P}-q/2)^2-m^2+{i\over 2}(\sigma G)}\Gamma_\nu\,\times
\label{eq:A.3}
\end{equation}
$$(\hat{\cal P}+\hat q/2){1\over ({\cal P}+q/2)^2-m^2+{i\over 2}\sigma G}\}
{}.
$$
The mass terms  drop out because of the chiral structure of
$\Gamma$ matrices.
\par For the magnetic emission we merely expand
the denominators of the quark propagators in $\sigma G$ keeping only
the linear term in the expansion. Since we are aimed at linear terms only,
in all other places one can  immediately substitute $P\rightarrow p$,
the normal momentum, to be treated as a regular integration variable.
Thus, calculation of this part is trivial. After some straightforward algebra
we get in this way
\begin{equation}
{\rm Im} (D^a_{\mu\nu})_{\rm magn}=-4i\{G^a_{\mu\alpha}q^\alpha q_\nu
+(\mu\leftrightarrow\nu) +i(\tilde{G}^a_{\mu\alpha}q^\alpha q_\nu
+(\mu\leftrightarrow\nu))\}\,\times
\label{eq:A4}
\end{equation}
$$
\{{1\over 3q^2}[{3-v^2\over 2v^2}(v\partial v) -1]{\rm Im}D_0\}.
$$
Here $D_0$ is the bare loop of scalar colorless
particles:
\begin{equation}
D_0=-i\int {d^4p\over (2\pi)^4} {1\over (p+q/2)^2-m^2}\,\times\,
{1\over (p-q/2)^2-m^2}
,
\label{eq: A.5}
\end{equation}
$$
{\rm Im}D_0={v\over 16\pi},\quad v=(1-4m^2/q^2)^{1/2}.
$$
\par Situation with the charge emission is somewhat more complicated.
 Now we
omit $(\sigma G)$ terms in eq. (\ref{eq:A.3}). Then taking the trace over the
Lorents ($\gamma$ matrix) indices and isolating the structure of interest
we arrive at
\begin{equation}
(D_{\mu\nu})_{\rm charge}=8i{\rm Tr}\,\{t^a ({\cal P}-q/2)_\mu
{1\over ({\cal P}-q/2)^2-m^2}({\cal P}+q/2)_\nu
{1\over ({\cal P}+q/2)^2-m^2}
\label{eq:A.6}
\end{equation}
$$+(\nu\leftrightarrow\mu
)\}$$
It is worth emphasizing that eq. (\ref{eq:A.6}) is automatically p-even
(there is no p-odd contribution from the charge emission).
It produces the structure
$(D^a_{\mu\nu})_{\rm charge}=F_{\rm ch}(G_{\mu\alpha}^a q^\alpha q_\nu
+G^a_{\nu\alpha}q^\alpha q_\mu )$.
In order to determine the coefficient $F_{\rm ch}$ in front of this structure
it is convenient to apply the following trick. Let us introduce an arbitrary
 vector $n_\alpha$ with the properties $( nq)=0, n_\alpha G_{\alpha\beta}\ne
0$.
Then it is easy to obtain
$$
F_{\rm ch}(nG^aq)=4i{\rm Tr}\,\{t^a ({\cal P}n)\times
$$
\begin{equation}
\left[ {1\over ({\cal P}-q/2)^2-m^2}{1\over ({\cal P}+q/2)^2-m^2}
-{1\over ({\cal P}+q/2)^2-m^2}{1\over ({\cal P}-q/2)^2-m^2} \right]\} .
\label{eq:A7}
\end{equation}
In eq. (\ref{eq:A7}) we omitted some tadpole pieces that have no imaginary
parts.
\par Let us take into account now that for any 2 operators
\begin{equation}
{1\over O_1}{1\over O_2}-{1\over O_2}{1\over O_1}=
{1\over O_1}{1\over O_2}(O_1O_2-O_2O_1){1\over O_2}{1\over O_1}.
\label{eq:A.8}
\end{equation}
This fact implies that in the square brackets in eq. (\ref{eq:A7})
we get the commutator
\begin{equation}
[(({\cal P}-q/2)^2-m^2), (({\cal P}+q/2)^2-m^2)]
=4i{\cal P}^\alpha G_{\alpha\beta}q^\beta .
\label{eq:A9}
\end{equation}
Now $G_{\alpha\beta}$ is explicitly singled out, and we can substitute
${\cal P}\rightarrow p$ everywhere.
As a result
\begin{equation}
F_{\rm ch}(nG^aq)=-8G^a_{\alpha\beta}\int {d^4p\over (2\pi)^4}
p^\rho p^\alpha{1\over ((p+q/2)^2-m^2)^2}{1\over ((p-q/2)^2-m^2)^2}n_\rho
 q^\beta
\label{eq:A.10}
\end{equation}
The integral in eq. (\ref{eq:A.10}) can be done directly. Alternatively,
one can reduce its imaginary part to ${\rm Im} D_0$ by differentiating
$D_0$ with respect to $q^2$ and $m^2$. Both derivatives are expressible in
terms of $\partial /\partial v$. Moreover, since ${\rm Im} D_0$ is linear
in $v$ the second derivative $\partial^2 /\partial v^2 {\rm Im} D_0$
vanishes.
In this way we arrive at
\begin{equation}
({\rm Im} D^a_{\mu\nu})_{\rm charge}=-8i{1\over 3q^2}({v^2-3\over 4v^2}
v{\partial\over\partial v}+{1\over 2}){\rm Im}D_0
(G_{\mu\alpha}^aq^\alpha q_\nu +(\mu\leftrightarrow\nu ))
\label{eq:A.11}
\end{equation}
$$
=-{2i(v^2-1)\over q^2v^2}{\rm Im}D_0\{G^a_{\mu\alpha}q^\alpha q_\nu
+(\mu\leftrightarrow\nu )\} .
$$
In the sum corresponding to both magnetic plus electric emissions the
p-even part cancels, and we are left with the p-odd part only,
\begin{equation}
{\rm Im}D^a_{\mu\nu}=2{1-v^2\over q^2v^2}{\rm Im}D_0\{\tilde G^a_{\mu\alpha}
q^\alpha q_\nu +(\mu\leftrightarrow\nu )\}
\label{eq:A.12}
\end{equation}
$$
={1\over 8\pi}{1-v^2\over q^2v}\{\tilde G^a_{\mu\alpha}q^\alpha q_\nu
+(\mu\leftrightarrow\nu )\} .
$$
This expression, obviously, is due to parity violating combination
of $\Gamma 's$ $(\gamma_\mu\gamma_5\times \gamma_\nu
+\gamma_\mu\times\gamma_\nu\gamma_5)$ in eq. (\ref{eq:A.1}).
\par It is instructive to notice that at $m^2\rightarrow 0$
${\rm Im} D^a_{\mu\nu}$ tends to a delta function
\begin{equation}
{\rm Im} D^a_{\mu\nu}\mid_{m^2\rightarrow 0}\rightarrow
{1\over 4\pi}\delta (q^2)\{\tilde G^a_{\mu\alpha}q^\alpha q_\nu
+(\mu\leftrightarrow \nu)\} .
\label{eq:A.13}
\end{equation}
\par Let us proceed now  to calculation of the full imaginary part
in eq. (\ref{eq:7}). This calculation reduces now to  (see Fig. 1)
\begin{equation}
{\rm Im} {\cal M}\equiv {\rm Im}\int d^4x iT\{O_2 (x)\tilde O_1^+(0)\}=
\label{eq:A.14}
\end{equation}
$$
=2\int {d^3\vec p\over 2p^0(2\pi )^3}{\rm Im}D^a_{\mu\nu}
(\bar b t^a\Gamma^\mu\hat p\Gamma^\nu b)
$$
where $p$ is the momentum of the massless $s$ quark
($p$ is on mass shell, $p^2=0$). The factor 2 in the r.h.s.
reflects the factor of 2 in the definition of $\tilde O_1$.
\par Integration over $p$ is best done in the rest frame of $b$.
Integrating over angles one finds
\begin{equation}
{\rm Im}{\cal M} ={1\over 24\pi^3}\int {m^2ds\over s^2v}{1\over M^3}
(\bar b \tilde G_{0i}^at^a\gamma^i\gamma^5b)(M^2-s)^2(2M^2+s)
\label{eq:A.15}
\end{equation}
where $s\equiv q^2$ is the invariant mass of the $c\bar c$ pair,
 $M$ is the b-quark mass.
\par Doing the integral in eq. (\ref{eq:A.15}) we arrive to eqs.
(\ref{eq:11}), (\ref{eq:12}).
\par It is instructive to check that the gluon emission from the line
lying outside the closed quark loop vanishes, irrespectively of the relation
 between the masses of the quarks that propagate along the
  different lines.
To this end it is sufficient to use eq. (\ref{eq:6}) for the quark Green
function. The quark loop is formed from two bare propagators, while in
the line lying outside the loop we take the $O(G)$
term in eq. (\ref{eq:6}). The terms with the even number of
$\gamma$ matrices
 (omitted in eq.
(\ref{eq:6}) ) are irrelevant, since they drop out due to
the chiral structure of $\Gamma$'s.
Using eq. (\ref{eq:6}) it is easy to see that the closed quark loop
yields
$(2x_\mu x_\nu-x^2g_{\mu\nu})$
 times a function of $x^2$. The line lying outside the
closed quark loop produces $x^\alpha\tilde G_{\alpha\varphi}
\bar b \Gamma^\nu\gamma^\varphi\gamma^5\Gamma^\mu b$ times a function of
$q^2$. Convoluting these two structures we get identical zero.
\par Our last explanatory remark refers to
eq. (\ref{eq:5}). Calculating the bare loop of Fig. 7 we get a local
operator of the type
$\bar b\partial^4\hat \partial b$, the intermediate quark line
shrinks into a point. The operator
$\bar b\partial^4\hat \partial b$ contains derivatives which
 should be rewritten in terms of covariant
quantities, say, for the purpose of using the equations of
motion.
\par Let us assume for definiteness that
the operator $\bar b\partial^4\hat \partial b$
is taken at the origin (this assumption is technical and final result
does not depend on it). The four-potential can be chosen in the form
\begin{equation}
A^a_\mu={1\over 2}x^\rho G_{\rho\alpha} .
\label{eq:A.16}
\end{equation}
The fact that $A(0)=0$ is the consequence of the Schwinger gauge condition
imposed on the external field.
Eq. (\ref{eq:A.16})
 assumes this condition, and,  if not stated to the contrary,
all specific calculations made above are done in this gauge. In particular,
the assertion that there is no $A^a_\mu$ leg associated with the line lying
outside the closed quark loop is also specific to this gauge.
In the Schwinger gauge the result for the interference term $<O_2O_1^+>$
reduces to $<O_2\tilde O_1^+>\sim \tilde G^a_{\mu\nu}$
plus the correlator ${1\over N_c}<O_2,O_2^+>$ given by the only graph
depicted on Fig. 7.
(modulo  terms of higher order in $A^a_\mu$).
Using eq. (\ref{eq:A.16}) and the remarks above  it is easy to obtain the
 expression
for $P^4\hat P$. We start from
\begin{equation}
\bar b (0) {\cal P}^4\hat{\cal P} b(0)=\bar b(0) (P+A)^4(\hat P+\hat A)b(0),
P=i\partial_\mu
\label{eq:A.17}
\end{equation}
and commute all the $A$'s to the left of all $P$'s. Once $A$ is in the
leftmost
position we use the fact that $A(0)=0$.
In the Schwinger gauge
$$[P_\beta A_\alpha]={i\over 2}G_{\beta\alpha} .$$
In this way we arrive at
\begin{equation}
{\cal P}^4\hat{\cal P}=P^4\hat P-{i\over 2}[(G\sigma)\hat P^3
-\hat P^3 (G\sigma)] .
\label{eq:A.18}
\end{equation}
Besides that
\begin{equation}
{\cal P}^2=\hat{\cal P}^2-{i\over 2}(G\sigma) .
\label{eq:A.19}
\end{equation}
Combining equations
(\ref{eq:A.18} ) and (\ref{eq:A.19}) one finds
\begin{equation}
P^4\hat P=\hat{\cal P}^5-{i\over 2}[(G\sigma)\hat P^3
+\hat P^3 (G\sigma)]+O(G^2)
\label{eq:A.20}
\end{equation}
which is equivalent to eq. (\ref{eq:5}).

\newpage
{\bf Figure Captions}
\vskip0.8in
{\bf Fig.1} Imaginary part of such graphs determines the inclusive widths
in the asymptotic regime.

{\bf Fig.2} A subgraph that determines the invariant mass distribution.

{\bf Fig.3} Creation of the states of the type (i) in the color transparency
picture.

{\bf Fig.4} Creation of the states by the operator $O_1$ at the
level ${1\over N_c}$.

{\bf Fig.5} Creation of the states by the operator $O_2$ at the level
${1\over N_c}$.

{\bf Fig.6} Calculation of the correlator ${1\over N_c}<O_2,O_2^+>$.

\newpage
\begin{thebibliography}{10}

\bibitem{BUV} I. Bigi, N. Uraltsev and  A. Vainshtein, preprint FERMILABPUB-
92/158-T.

\bibitem{BS1}
B. Blok and M. Shifman, preprint NSF-ITP-92-103.

\bibitem{BS2}
B. Blok and M. Shifman, preprint NSF-ITP-92-76 (Nuclear
Physics B, in press).

\bibitem{BSW}
 M. Bauer, B. Stech and M. Wirbel, Z. Phys. C34 (1987) 103.

\bibitem{BGR}
A.J. Buras, J.-M. Gerard and R. Rueckl, Nucl. Phys. B268 (1986) 16.

\bibitem{XX}
 J. L. Cortes, X.Y. Pham and A. Tounsi, Phys. Rev., D25 (1982) 188.

\bibitem{N}
V.  Novikov, M. Shifman, A. Vainshtein and V. Zakharov,
Fortsch. Phys. 32 (1984) 585.

\bibitem{S}
 M. Shifman, preprint TPI-MINN-91/46-T (Nuclear Physics B, in press).

\bibitem{VS}
 M.Voloshin and  M. Shifman,
Yad. Fiz. 47 (1988) 80 [Sov. Journ. Nucl. Phys. 47 (1988) 511].

\bibitem{B}
J. Bjorken, preprint SLAC-PUB-5278 (1990), Invited talk at Les Recontre
de la Valle d'Aosta, La Thuille, Italy (1990)

\bibitem{DZ}
A. Dolgov and V. Zakharov, Nucl. Phys. B27 (1971) 525.

\bibitem{T}
G. 't Hooft, in: {\em Recent Developments in Gauge Theories},
eds. G. 't Hooft et al.l (Plenum Press, New York, 1980), page 241.

\bibitem{L}
 G. Bodwin, E. Braaten, T.C. Yuan and G. P. Lepage,
preprint ANL-HEP-PR-92-63, NUHEP-TH-92-16.

\end{thebibliography}
\end{document}